\documentclass[11pt]{letter}
\usepackage{amsmath}
\usepackage{amssymb}
\usepackage[dvips]{graphicx}
\DeclareFontFamily{U}{msb}{}
\DeclareFontShape{U}{msb}{m}{n}{ <5> <6> <7> <8> <9> gen * msbm
        <10> <10.95> <12> <14.4> <17.28> <20.74> <24.88> msbm10}{}
\DeclareSymbolFont{AMSb}{U}{msb}{m}{n}
\DeclareMathSymbol{\realset}{\mathalpha}{AMSb}{"52}

\newcommand{\osd}[1]{\overset{\scriptscriptstyle #1}{)}}
\newcommand{\ose}[1]{\overset{\scriptscriptstyle #1}{(}}

\begin{document}

\begin{center}
\Large 
{\bf A Central Partition of Molecular Conformational Space. \newline
     IV. Extracting information from the graph of cells.}
\end{center}

\vspace*{6mm}
\begin{center}
{\Large Jacques Gabarro-Arpa}
\end{center}

\vspace*{6mm}
\hspace*{26mm}
Ecole Normale Sup\'erieure de Cachan, LBPA,CNRS UMR 8113      \newline
\hspace*{26mm}
61, Avenue du Pr\'esident Wilson, 94235 Cachan cedex, France  \newline

\noindent \hspace*{60mm} Email: jga@gtran.org

\vspace*{4mm}
In previous works (Gabarro-Arpa, J. Math. Chem. 42 (2006) 691-706) a procedure was decribed for dividing the $3 \times N$-dimensional conformational space of a molecular system into a number of discrete cells, this partition allowed the building of a combinatorial structure from data sampled in molecular dynamics trajectories: the graph of cells or $\mathbf{G}$, that encodes the set of cells in conformational space that are visited by the system in its thermal wandering.
Here we outline a set of procedures for extracting useful information from this structure: 1st) interesting regions in the  volume occupied by the system in conformational space can be bounded by a polyhedral cone, whose faces are determined empirically from a set of relations between the coordinates of the molecule, 2nd) it is also shown that this cone can be decomposed into a set of smaller cones, 3rd) the set of cells in a cone can be encoded by a simple combinatorial sequence.

\vspace*{2mm}
\hspace*{4mm} {\it Keywords}
{\bf Molecular Conformational Space, Hyperplane Arrangement, Face Lattice, Molecular Dynamics} 

{\it Mathematics Subject Classification: } 52B11, 52B40, 65Z05

{\it PACS:} 02.70.Ns

\newpage
{\bf 1. Introduction}

The aim of this series of papers [1-4] is to build a set of mathematical tools for studying the energy landscape of proteins [5,6,7], and the present paper is a step further towards this goal.

The energy surface of proteins is the essential tool for understanding the physico-chemistry of basic biological processes like catalysis [7]. It is also a complex multidimensional structure that can only be built from the knowledge of the complete dynamical history of the molecule, which is currently out of reach for conventional molecular-dynamics simulations (thereafter referred as MDS) [7]. One reason is that in an MDS trajectory the position of every atom in the molecule is calculated with an accuracy of a hundredth of angstr\"{o}m, which quickly overhelms even the most powerful computers.
The approach taken here consists in encoding the small movements of a molecular system by means of some combinatorial structure, that allows to generate the set of realizable combinations of these movements.

Within this approach, the $3D$-structures of protein molecules are encoded into binary objects called \textbf{dominance partition sequences} (DPS) [1-4], these are the generalization of a combinatorial structure known as noncrossing partition sequences [8]. In this context the basic structure for studying the molecular dynamics is the set of $3D$-conformations that have the same DPS, these form a connected region in molecular conformational space\footnote[1]{For an $N$-atom molecule it is a $3 \times N$-dimensional space where each point corresponds to a $3D$ molecular conformation.}
(in what follows abridged to $CS$) called \textbf{cell}, thus DPSs generate a partition of $CS$ into disjoint cells. Partitions are a useful tool for studying multi-dimensional spaces, in our case they systematically spann a much wider volume range than the set of points along a random trajectory curve generated by a MDS, they have also been used in many other contexts [5,6,9].

The aim of the preeceding papers [1-4] was to construct a graph whose nodes are the cells visited by the molecular system in its thermal wandering, two important properties of partition sequences make this construction possible :

\begin{enumerate}
 \item DPSs are \textbf{hierarchical} structures: partition sequences encoding different sets of cells can be merged into a 
       new partition sequence encoding the union set, and the process can be repeated with the new sets of cells, thus creating a
       hierarchy.
       The importance of this property is that climbing the hierarchy ladder \textit{the number of cells increases exponentially
       while the sequence length increases only linearly}.
       This compact coding makes possible the construction of a graph representing huge regions of $CS$ whose size does not exceed the
       memory of a workstation computer, while keeping at the same time the essential information about the molecular structures.
 \item DPSs are \textbf{modular} structures: partition sequences can be decomposed into subsequences that are embedded in different
       conformational subspaces. This allows to define a \textbf{composition law}: if two partition sequences from two different
       subspaces share the same sequence for the intersection subspace, then joining both sequences gives a realizable
       sequence\footnote[2]{That corresponds to an existing set of cells.} [4].
\end{enumerate}

The first property tells us that the graph can be constructed, the second suggests how to build it: a molecular structure can be decomposed into sets of four atoms, its smallest $3D$ components, by composing the graphs of these one can build the graph of the molecule.

Atoms in MDSs are represented as pointlike structures surrounded by a force field [10,11], the convex enveloppe of a set of 4 points in $3D$-space is an irregular polytope called a \textbf{4-simplex} 
or \textbf{simplex}\footnote[3]{In what follows this denomination will be used to designate ordered sets of 4-atoms/points.}.
The conformational space of these sets is relatively small with 13824 cells, of these only a fraction is visited by the system. With a $CS$ so small it can be plausibly assumed that the accessible cells are all visited during a MDS run.

The method for building the graph that was proposed in [2] consists in 
\begin{enumerate}
 \item Establishing a morphological classification of simplexes, where each class is defined by a set of geometrical constraints.
 \item The geometrical constraints that define a class allow to calculate the set of accessible  cells in a simplex $CS$ [4],
       thus to each class we can associate a graph where the nodes are the cells from this set with edges towards adjacent cells.
 \item On the other hand computer simulations of protein dynamics show [2,4] that in a protein structure the majority of simplexes
       evolve within a reduced number of morphologies. For each 4-atom set in the molecule the graph of its $CS$ is built by merging
       the graphs of the visited simplex morphologies.
 \item The $CS$ graph of the molecule, that was called the \textbf{graph of cells} or $\mathbf{G}$ in [4], can be built by composing
       the $CS$ graphs of the different simplexes.
\end{enumerate}

The graph of cells allows to enumerate exactly the set of visited cells in conformational space, but since the cells are encoded in a compact form unwrapping them completely is probably algorithmically hopeless. Instead here we propose the construction of more manageable coarse-grained encodings that, using the information from $\mathbf{G}$, can be recursively decomposed into progressively  fine-grained ones. This subject is developped in the next five sections:

\begin{itemize}

 \item Section 2 is a graph of cells oriented description of the basic mathematical framework.
 \item Section 3 is about the basic mathematical properties of $\mathbf{G}$.
 \item Section 4 describes how to determine, from empirical data, a conical boundary for the region occupied by the system in $CS$.
 \item Section 5 shows how to decompose this cone boundary into a set of smaller cones.
 \item Section 6 is devoted to describing a combinatorial sequence that encodes the conical boundary in its most compact form.

\end{itemize}

\vspace*{4mm}
{\bf 2. The basic construction}

It was shown [1] that the conformational space of a molecule of $N\!+\!1$ atoms $\realset^{3\!\times\!N}$\footnote[4]{$N\!+\!1$ is because the translation symmetry makes one dimension spurious [1,4].}
could be described to a fair degree of accuracy by means of the partition generated by a set of hyperplanes passing through the origin that form a Coxeter reflection arrangement\footnote[5]{So called because a reflexion through one of the hyperplanes leaves the arrangement unchanged.} denominated $\mathcal{A}^{N}$ [8,12], moreover the reflections form a symmetry group that is isomorphic to the symmetric group.

In our description of $CS$ we have three independent arrangements one for each coordinate $(x, y, z)$, i.e. $\mathcal{A}^{3\!\times\!N} = \mathcal{A}^{N}\!\times\!\mathcal{A}^{N}\!\times\!\mathcal{A}^{N}$, that generate three partitions of $\realset^{3\!\times\!N}$, each dividing $\realset^N$ into a hierarchical set of regions shaped as polyhedral cones denominated \textbf{cells}.
The hyperplanes in our partition are defined as

$\mathcal{H}_{ij}: x_{i}\!-\!x_{j}=0 \ \ \ , \ \ \ 1 \leq i < j \leq N\!+\!1$ \hspace*{20mm} (1)

each $\mathcal{H}_{ij}$ divides $\realset^{N}$ into three regions : 

$x_{i} < x_{j}$ \ \ , \ \ $x_{i} = x_{j}$ \ \ and \ \ $x_{i} > x_{j}$ \hspace*{26mm} (2)

in the first case we say that $x_{j}$ \textbf{dominates} $x_{i}$, in the second case neither $x_{i}$ nor $x_{j}$ dominates, in the last case $x_{i}$ dominates $x_{j}$.
As cells are bounded by the hyperplanes (1) a consequence of (2) is that the points inside a given cell (in $x$, $y$ or $z$) have the following property:

$x_{i_{1}} \leq x_{i_{2}} \leq x_{i_{3}} \leq . . . \leq x_{i_{N\!-\!2}} \leq x_{i_{N\!-\!1}} \leq x_{i_{N}}$ \hspace*{14.5mm} (3)

where the sequence $(i_{1}, i_{2}, i_{3}, ... i_{N\!-\!2}, i_{N\!-\!1}, i_{N})$ is a permutation of the set $\mathcal{Z}_{N\!+\!1} = (1, 2, 3, ... \ N, N\!+\!1)$, reflecting a point through $\mathcal{H}_{ij}$ is equivalent to permute the coordinates $i$ and $j$ [8]. Thus a cell where a strict "less than" relation holds for every pair of coordinates in (3) is encoded by the \textbf{dominance sequence}

$(i_{1})(i_{2})(i_{3}) ... (i_{N\!-\!2})(i_{N\!-\!1})(i_{N})$ \hspace*{37.2mm} (4a)

while for a cell where $x_{i_{\alpha}} = x_{i_{\alpha\!+\!1}} = ... = x_{i_{\alpha\!+\!r}}$, for $r\!+\!1$ consecutive indices $(i_{\alpha}, i_{\alpha\!+\!1}, ... i_{\alpha\!+\!r})$ in (3) will be encoded by the dominance sequence

$(i_{1})(i_{2})(i_{3}) ... (i_{\alpha} i_{\alpha\!+\!1} ... i_{\alpha\!+\!r}) ... (i_{N\!-\!1})(i_{N})(i_{N\!+\!1})$ \hspace*{12mm} (4b)

the first (4a) represents an $N$-dimensional cell while (4b) is a $(N\!-\!r)$-dimensional cell because it corresponds to the intersection of the hyperplanes $\mathcal{H}_{ij}$ with $i , j \in (i_{\alpha} i_{\alpha\!+\!1} ... i_{\alpha\!+\!r})$.

{\bf Definition 1}. {\it The \textbf{position} of a coordinate $x^{c}_{i}$ in a cell of dimension $N$ is the position of the index $i$ in the dominance sequence of $c$}.

An alternative encoding of cells is by means of an $N\!\times\!N$ antisymmetric {\bf sign matrix} $\mathcal{S}^{c}$, where $c$ stands for $x$, $y$ or $z$. Let $1 \leq i < j \leq N\!+\!1$, then for an arbitrary point $x$ the matrix elements $\mathcal{S}^{c}$ for the $c$ coordinates are defined:

$\mathcal{S}^{c}_{ij} = -$   if $x^{c}_{i} < x^{c}_{j}$ \\
$\mathcal{S}^{c}_{ij} = 0$ \ if $x^{c}_{i} = x^{c}_{j}$ \hspace*{55mm} (5) \\
$\mathcal{S}^{c}_{ij} = +$   if $x^{c}_{i} > x^{c}_{j}$

As it was explained in [1,4] a direct consequence of (3) is that $\mathcal{S}^{c}$ can be interpreted as the incidence matrix of a digraph with no directed cycles, and the cell encodings (3) and (5) can be readily interconverted into one another 

{\bf Lemma 1}. {\it Contiguous cells in space have different dimensionalities}.

Crossing to a contiguous cell implies going between two regions in (2), so one element $\mathcal{S}^{c}_{ij}$ in (5) changes its value, and this change can never be between $+$ and $-$ because this would mean crossing $\mathcal{H}^{c}_{ij}$ avoiding the region $c_{i} = c_{j}$.

{\bf Definition 2}. {\it A \textbf{contiguous set} are all the $n$-dimensional cells contiguous to a $(n\!-\!1)$-dimensional \textbf{separator cell}}.

This allows to build a hierarchical structure: the \textbf{cell lattice poset}, that results from ordering contiguous cells by dimensionality [1,13].

Consider two arbitrary subpartitions $\mathcal{A}_{a}^{d_{a}}$ and $\mathcal{A}_{b}^{d_{b}}$ of $\mathcal{A}^{N}$, corresponding to the sets of indices $\chi_{a} = (i_{a_{1}}, i_{a_{2}}, ... i_{a_{d_{a}+1}}) \subset \mathcal{Z}^{d_{a}\!+\!1}$ and $\chi_{b} = (i_{b_{1}}, i_{b_{2}}, ... i_{b_{d_{b}\!+\!1}}) \subset \mathcal{Z}^{d_{b}\!+\!1}$ respectively, and let $\chi_{a \cap b} = \chi_{a} \cap \chi_{b}$ be the set of indices that are common to both partitions.

{\bf Definition 3}. {\it Two cells $\zeta_{a} \in \mathcal{A}_{a}^{d_{a}}$ and $\zeta_{b} \in \mathcal{A}_{b}^{d_{b}}$ with sign matrices $\mathcal{S}^{a}$ and $\mathcal{S}^{b}$ respectively, are said to be \textbf{compatible} if $\mathcal{S}^{a}_{ij} = \mathcal{S}^{b}_{ij} \ \ \forall \ i, j \in \chi_{a \cap b}$}.

{\bf Lemma 2}. {\it The cell $\zeta_{a} \in \mathcal{A}_{a}^{d_{a}}$ is the projection of all the cells in $\mathcal{A}^{N}$ whose sign matrix $\mathcal{S}$ is such that $\mathcal{S}_{ij} = \mathcal{S}^{a}_{ij} \ \ \forall \ i, j \in \chi_{a}$}.

This is an inmediate consequence of (3) and (5).

Let $\Xi_{a}$ and $\Xi_{b}$ be the set of cells in $\mathcal{A}^{N}$ that are projected on $\zeta_{a}$ and $\zeta_{b}$ respectively

{\bf Lemma 3}. {\it The set $\Xi_{a} \cap \Xi_{b}$ is non empty iff $\zeta_{a}$ and $\zeta_{b}$ are compatible}.

Suppose we have $\xi \in \Xi_{a}$ but $\xi \not\in \Xi_{b}$, this means that the relative positions of the set of indices $\chi_{b \setminus a} = \chi_{b} \setminus \chi_{a}$ in the dominance sequence (4) is not the same as in $\zeta_{b}$, since the reflexion group of the arrangement is the symmetric group there always will be a set of permutations/reflections that sorts the indices $\chi_{b \setminus a}$ in the dominance sequence in the same order as in $\zeta_{b}$, this generates a cell $\xi^{'} \in \Xi_{a} \cap \Xi_{b}$.

\vspace*{4mm}
{\bf 3. The graph of cells}

Lemmas 2 and 3 suggest that $\mathcal{A}^{3\!\times\!N}$ can be built by merging partitions of lower dimensionality. The smallest $3D$ system is a set of 4 atoms, and $\mathcal{A}^{3\!\times\!4 - 1}$, the partition of its $CS$, has exactly 13824 cells, a computational complexity within the range of a desktop computer. Moreover, as stated in the introduction it can be reasonably assumed that such small $CS$ can be thoroughly scanned by a MDS.

Following the procedure proposed in refs. [2,3,4] (outlined in the introduction) we can build the $CS$ of a molecular system from the $CS$ of the simplexes. For this, we need to construct the graph of cells or $\mathbf{G}$ which is defined as follows:

{\bf Definition 4}. \textit{Two simplexes are \textbf{adjacent} if they share a face}.

{\bf Definition 5}. \textit{The nodes of $\mathbf{G}$ are the visited cells of each simplex with edges towards the compatible cells  of adjacent simplexes}.

{\bf Definition 6} \textit{A \textbf{tranversal} is a subgraph of $\mathbf{G}$ whith nodes exactly one cell from every simplex such that every two cells from adjacent simplexes are compatible}.

$\mathbf{G}$ embodies all the information contained in the $CS$ of a molecular system since

{\bf Theorem 1}. {\it The cells in a tranversal are the projections of a single cell in $CS$}

By lemma 3 the cells in the transversal are the projection of at least one cell in $\mathcal{A}^{3\!\times\!N}$, that cell is unique because if there were two, for instance, their sign matrices would not be the same, say that the element $\mathcal{S}^{c}_{ij}$ is different, then there is a set of $\binom{N-1}{2}$ simplexes that harbor the indices $i$ and $j$ and within this set each simplex is adjacent to $2\times(N-3)$ other simplexes, from definition 5 adjacent simplexes have to be compatible and the element $ij$ in their sign matrix must be the same for all, invalidating our assumption.

{\bf Corollary}. {\it In $\mathbf{G}$ a node that fails to form an edge with an adjacent simplex cannot exist since it is geometrically inconsistent}.

A useful structure derived from $\mathbf{G}$ is its compact form $\mathbf{C}$ obtained by recursively substituing every contiguous set of $n$-dimensional nodes by their $(n\!-\!1)$-dimensional separator cell.

Finally a cell from $\mathcal{A}^{3\!\times\!N}$ is a class in an equivalence relation, since it contains all the $3D$-structures that have the same dominance sequence. In what follows we use the terms cell and $3D$-structure interchangeably.

{\vspace*{4mm}
{\bf 4. Determining a conical boundary for the molecular dynamics trajectory}

$\mathbf{G}$ is a huge structure and it is probably useless to try to explore it in full, rather the approach we take here is how to focus on regions (subgraphs) where we can expect to extract useful information. We start with the problem of finding the bounds of interesting regions, with a concrete exemple concerning a 2.1 ns pancreatic trypsin inhibitor (PTI) [14] MDS that was fully described in [15].

As in [15] we restrict ourselves to study the motion of $C^{n}_{\alpha}$ carbons each bearing a number $n$ that reflects the linear order of residues along the polypeptide chain, as our description of $CS$ is strictly modular any conclusion that can be drawn on any subset of atoms is automatically valid for the whole structure.

An information easily extracted from a MDS are the dominance relations matrices $DR^{c}$, where $c$ stands for either $x$, $y$ or $z$, each element of these matrices defines the equation of a face in a polyhedral cone, it encloses the region that the molecular system occupies in $CS$. The determination of the $DR^{c}$s from the MDS [15] takes the following steps:
\begin{itemize}
\item First, the simplex corresponding to the residue numbers $\mathsf{S}_{r} = \{6,36,40,47\}$ was selected as the reference simplex because all along the MDS it stays within one morphological class, and because it spans a wide volume across the molecule.
\item Second, the coordinates of $\mathsf{S}_{r}$ in the $1^{st}$ MD frame were taken as a reference and the other frames were rotated and translated so that the RMS between $\mathsf{S}_{r}(1)$ and $\mathsf{S}_{r}(f)$ be a minimum [16].
\item Third, the quantities $DR^{c}_{ij} \ , \ 1 \leq i < j \leq N\!+\!1$, were determined 
      \begin{itemize}
      \item $DR^{c}_{ij} = +$ , $DR^{c}_{ji} = -$ if $C^{i}_{\alpha_{c}} > C^{j}_{\alpha_{c}}$ for all coordinate frames.
      \item $DR^{c}_{ij} = -$ , $DR^{c}_{ji} = +$ if $C^{i}_{\alpha_{c}} < C^{j}_{\alpha_{c}}$ for all coordinate frames.
      \item $DR^{c}_{ij} = DR^{c}_{ji} = 0$ if neither of the above relations holds. Also, by convention $DR^{c}_{ii} = 0$.
      \end{itemize}
\end{itemize}

The meaning of the matrix elements is obvious, if $DR^{c}_{ij} = +/-$ the trajectory always stays on the positive/negative side of $\mathcal{H}^{c}_{ij}$ (2), for $DR^{c}_{ij} = 0$ the trajectory can be on either side of $\mathcal{H}^{c}_{ij}$.
The matrices for $x$, $y$ and $z$ for the MDS [15] are shown in Figure 1, the number of non-zero terms in the matrix is the dimension of the cone.

{\bf Lemma 4}. {\it The  minimum position $\mathrm{min}^{c}_{\mu}$ of a coordinate $c_{\mu}$ is the number of matrix elements $DR^{c}_{\mu j} = +$ plus $1$ \ , \ $1 \leq j \leq N \ , \ j\!\neq\!\mu$, and the maximum position $\mathrm{max}^{c}_{\mu}$ is the minimum position plus the number of matrix elements $DR^{c}_{\mu j} = 0$ \ , \ $1 \leq j \leq N $}.

\vspace*{4mm}
{\bf 5. The fragmentation of the cone}

The dominance relations matrices $DR^{c}$ encode a lot of information about the structure of the volume occupied by the system in $CS$. They give us the range of positions of a given coordinate 

\newpage
\includegraphics{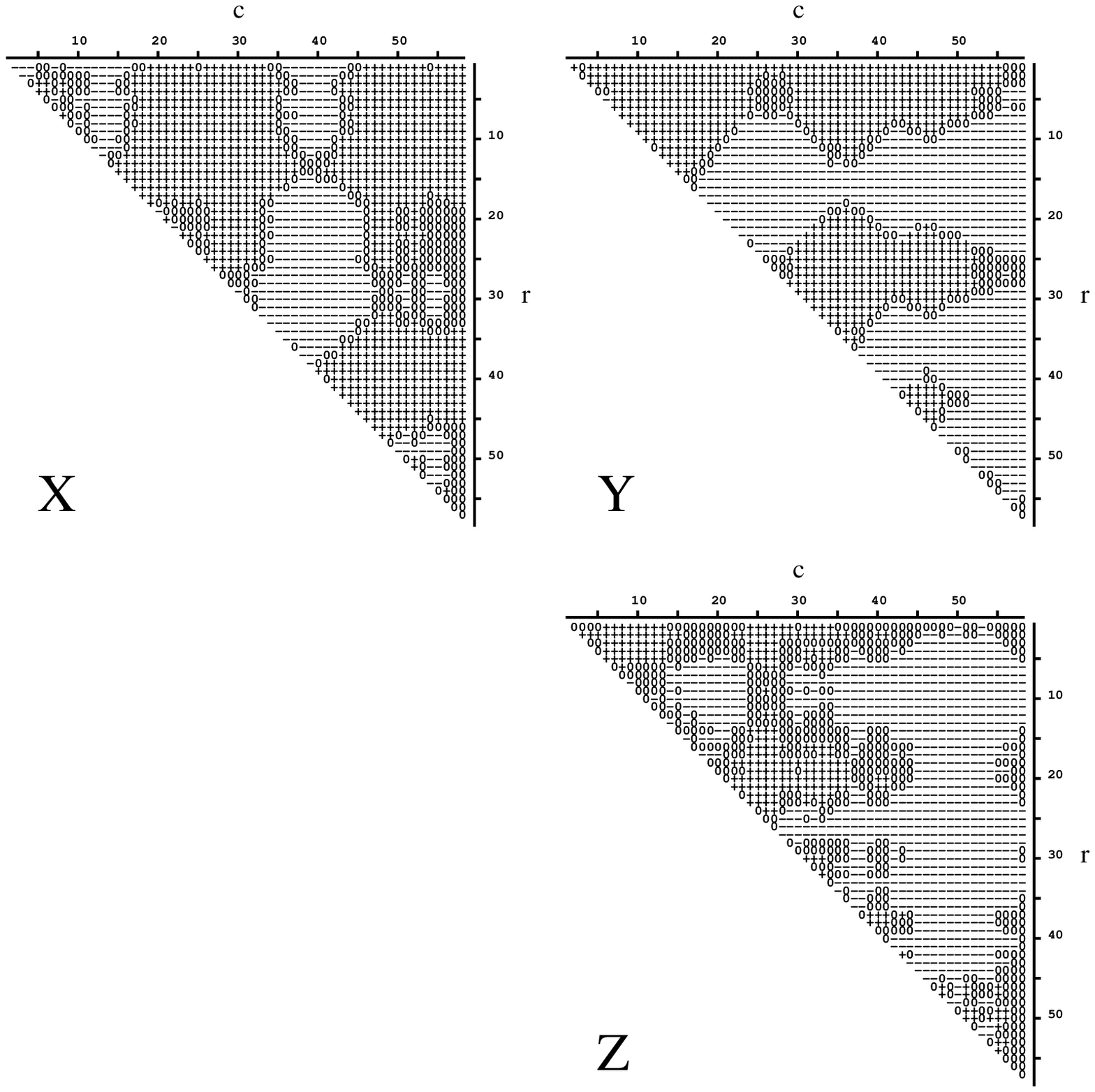}

\vspace*{4mm}
\begin{center}
\large{Figure 1} \\
\end{center}

\vspace*{4mm}
{\footnotesize
Antisymmetric dominance relations matrices for the $C_{\alpha}$ coordinates, only the upper triangle is shown.
For sake of clarity row and column amino acid numbers can be read from the annotated axes $r$ and $c$.
A matrix element can have three values
\begin{itemize}
\item[+] $x_{r} > x_{c}$ for all coordinate frames in the molecular dynamics run.
\item[-] $x_{r} < x_{c}$ for all coordinate frames in the molecular dynamics run.
\item[0] neither of the above relations holds.
\end{itemize}}

in the dominance sequence (3).

The index $\mu$ in the dominance sequence must always stay to the right of the elements it dominates if there are $n_{+}$ of such elements the minimum position of $\mu$ is $n_{+}\!+\!1$, on the other hand be $n_{0}$ the number of indiferent relations, $\mu$ can be either to the right or to the left of any of these then the maximum position of $\mu$ must be $n_{+}\!+\!n_{0}\!+\!1$.

A set of constraints can be defined for the cone

We can also extract from $DR^{c}$ sets of lower dimensional cells, these are useful for fragmenting $\mathbf{G}$ into subgraphs of more manageable size. To do this we can proceed as follows: we select indices ${\mu}$ and ${\nu}$ such that 

$\forall c \in \{x,y,z\} : \ \mathrm{max}^{c}_{\mu} > \mathrm{min}^{c}_{\nu} \ \ , \ \
                             \mathrm{max}^{c}_{\nu} > \mathrm{min}^{c}_{\mu}$ \ \ \ \ \ \ \ \ \ and \\
\hspace*{25mm} $\mathrm{MIN}(\mathrm{max}^{c}_{\mu},  \mathrm{max}^{c}_{\nu}) -
                \mathrm{MAX}(\mathrm{min}^{c}_{\mu},  \mathrm{min}^{c}_{\nu}) \geq h^c$ \hspace*{21.5mm} (6)

with \ \ $h^c = 1, \ 2, \ 1$ \ \ for \ \ $DR^{c}_{\mu\nu} = -1, \ 0, \ 1$ \ \ respectively.

we thus select pairs of atomic indices $\mu$ and $\nu$ whose ranges overlap in $x$, $y$ and $z$ simultaneously with intersection length $\geq ( h^x , h^y , h^z )$ respectively. For every pair index their ranges in any dimension are divided into three segments: left, middle (the intersection) and right; $\mu$, for instance, can occupy any position in the left and middle segments, while $\nu$ can be in the middle and right ones, this makes a total of 3 possibilities, 4 if $DR^{c} =0$ in which case $\mu$ and $\nu$ can be simultaneously in the middle segment. Obviously this can be extended to more than 2 indices: if $\mu\nu$, $\mu\omega$ and $\nu\omega$ have overlapping ranges, for instance, then there is a common overlapping range for $\mu$, $\nu$ and $\omega$ too, which in turn gives segmentation and occupation patterns for $\mu\nu\omega$.

\includegraphics{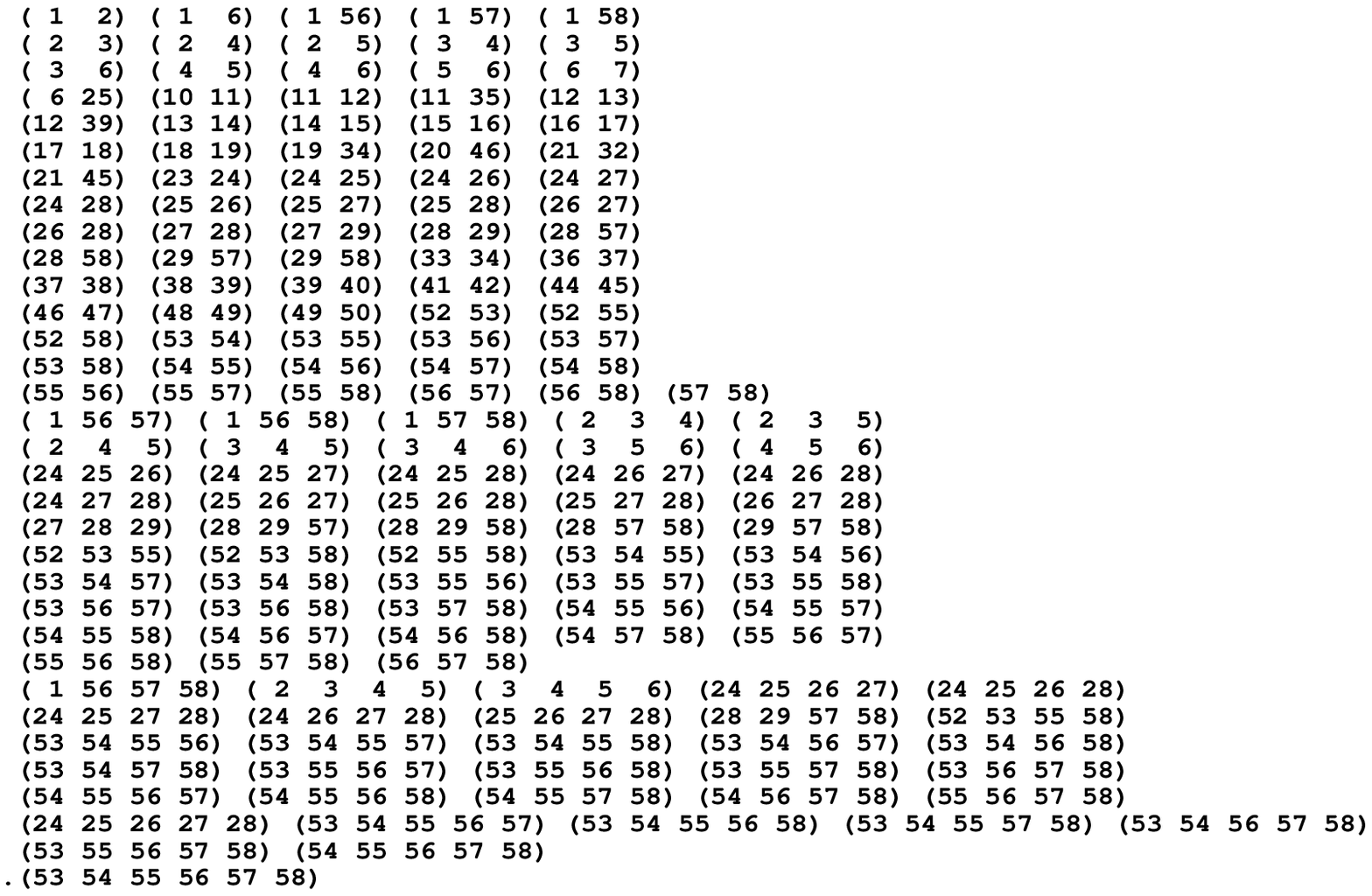}

\vspace*{6mm}
\begin{center}
\large{Table I} \\
\end{center}

{\footnotesize The complete sets of indices for the $\alpha$-carbons of the MDS described in [15] that conform to (6).}

The importance of overlapping indices is twofold:
\begin{enumerate}
\item A set of molecular conformational states can be determined from them using a minimum number of cells from $\mathbf{G}$:
      the indices being the same for $x$, $y$ and $z$ makes that occupation patterns for overlapping $\mu$ and $\nu$, for instance,
      can be deduced from the cells in $\mathbf{G}$ corresponding to the simplexes that bear these indices.
\item One can address the basic problem of how occupation states in the dominance sequence are correlated between different coordinates.
\end{enumerate}

The set of allowed overlapping indices than can be deduced from the $DR^{c}$ matrices in Figure 1 is given in table I.

This allows a procedure for fragmenting the cone in Figure 1 into smaller ones. From $\mathbf{G}$ we can deduce for each set of indices from Table I a number of local conformations, the valid combinations of these conformations will give us smaller cones whose cells have $DPS$s with mean positions $(x,y,z)$ much closer to the values of the cells in $\mathbf{G}$.

\vspace*{4mm}
{\bf 6. A combinatorial sequence for encoding cones.}

The codification of cones in conformational space could be much simplified by introducing a simple extension in the formalism  used for encoding dominance sequences : we allow expressions enclosed between parenthesis to overlap, and we distinguish between pairs of enclosing parenthesis by numbering them. Let us assume, for example, that we have a cone in $CS$ whith $DR$ matrix

$\begin{array}{rrrrrrr}
    & 1 & 3 & 4 & 7 & 8 & 9 \\
  1 &   & + & + & 0 & 0 & 0 \\
  3 & - &   & 0 & - & 0 & 0 \\
  4 & - & 0 &   & - & 0 & 0 \\
  7 & 0 & + & + &   & 0 & 0 \\
  8 & 0 & 0 & 0 & 0 &   & 0 \\
  9 & 0 & 0 & 0 & 0 & 0 &   \\
\end{array}. $  \hspace*{5.5mm} (7)

The sequence

$\ose{1} 3 \ 4 \
 \ose{2} 8 \ 9 \osd{1}  \
         1 \ 7 \osd{2}$ \hspace*{25mm} (8)

is meant to encode in one formula the sequences $(3 \ 4 \ 8 \ 9)(1 \ 7)$ and $(3 \ 4)(1 \ 7 \ 8 \ 9)$, these represent the totality of cells from $CS$ that lie inside the cone (7); structures like (8) will be designated as: {\bf generalized compact dominance sequences} ($GCDS$). Notice that parenthesis enclosed within parenthesis are not allowed within $GCDS$s since they are meaningless as dominance sequences.

$GCDS$s can encode huge numbers of cells from $CS$, for instance the first ten $\alpha$-carbons in our structure [14,15] evolve within a cone in $\mathcal{A}^{3\!\times\!10}$ encoded by the formula

$\{\{\ose{1}  1 \
     \ose{2}  5 \ 8 \
     \ose{3}  6 \osd{1} \
     \ose{4}  2 \
     \ose{5}  9 \osd{2} \
              7 \osd{3} \
             10 \osd{4} \
             3  \ 4 \osd{5} \}_{x},$ \\
\hspace*{0.8mm}
 $\{\ose{1} 10 \osd{1} \
    \ose{2}  9 \osd{2} \
    \ose{3}  8 \osd{3} \
    \ose{4}  7 \osd{4} \
    \ose{5}  5 \
    \ose{6}  4 \osd{5} \
             6 \osd{6} \
    \ose{7}  2 \
    \ose{8}  3 \osd{7} \
             1 \osd{8} \}_{y},$      \\
\hspace*{0.8mm}
 $\{\ose{1} 8 \
    \ose{2} 7 \ 10 \osd{1} \
            6 \  9 \osd{2} \
    \ose{3} 3 \  4 \ 5 \
    \ose{4} 1      \osd{3} \
            2      \osd{4} \}_{z}\}$ \hspace*{19.5mm}  (9)

that can be easily checked by comparing it with the $10\times 10$ upper-left submatrices in Fig. 1.

Not all the cones in $CS$ can be represented by $GCDS$s. A simple example will show us that the $x$-component of (9) can not be extended beyond the $14^{th}$ $C_\alpha$. Let

$\begin{array}{rrrrrrr}
    & 2 &              3  &              4  & 10 &             12  & 15 \\
  2 &   &              -  &              -  &  0 &              -  &  - \\
  3 & + &                 &              0  &  0 & \textcircled{-} &  0 \\
  4 & + &              0  &                 &  0 & \textcircled{-} &  0 \\
 10 & 0 &              0  &              0  &    &              0  &  0 \\
 12 & + & \textcircled{+} & \textcircled{+} &  0 &                 &  0 \\
 15 & + &              0  &              0  &  0 &              0  &    \\
\end{array}. $  \hspace*{22mm} (10)

be the $DR^{x}$ submatrix of the $\alpha$-carbons $2$, $3$, $4$, $10$, $12$ and $15$, it gives the sequence

$\ose{1}  2 \
 \ose{2} 10          \osd{1}  \
 \ose{3}  3 \ 4 \ 15 \osd{3}  \
         12          \osd{2}$ \hspace*{41.2mm} (11)

which is clearly inconsistent because $12$ dominates $3$ and $4$ but is on the same dominance level with $10$. One can perform slight modifications in (10) that transform (11) into a valid compact formula: setting to zero the circled components in (10) gives the $DR^{x}$ matrix of a $GCDS$-cone that encloses the cone (10). This modification allows us to extend the generalized dominance sequence $x$-component of (9) to the $20$ $\alpha$-carbons

$\{\ose{1}  19 \osd{1} \
   \ose{2}  20 \
   \ose{3}  18 \osd{2} \
   \ose{4}   1 \osd{3} \
            17 \
   \ose{5}   5 \
   \ose{6}   8 \
   \ose{7}   6 \osd{4} \
   \ose{8}   2 \
   \ose{9}   9 \osd{5} \
            11 \ 16 \osd{6} \
             7 \osd{7} \
   \ose{10} 10 \osd{8} \
             3 \  4 \osd{9} \
            15 \
   \ose{11} 12 \osd{10} \
   \ose{12} 14 \osd{11} \
            13 \osd{12}\}_{x}$ \hspace*{4.3mm} (12)

One can easily verify for every coordinate from Fig. 1 that the cone that bounds the evolution in $CS$ of every $10$ consecutive residues of the molecular structure is a $GCDS$-cone, for chains about $20$ residues and more this is generally no longer possible unless the value of some $DR^{c}$ elements are made zero as in (10). This result would seem to suggest that in the MDS from [15] thermodynamic equilibrium has not been attained, for instance in (10) $C_{\alpha_{10}}^{x}$ can swap dominance with $C_{\alpha_{3}}^{x}$, $C_{\alpha_{4}}^{x}$, $C_{\alpha_{7}}^{x}$ and $C_{\alpha_{15}}^{x}$, but pairs of cells with conformations where $C_{\alpha_{3}}^{x}$, $C_{\alpha_{4}}^{x}$ and $C_{\alpha_{7}}^{x}$ cross one another on the $x$-axis have not been visited by the MDS. This example clearly shows that $GCDS$-cones not only have a simple elegant formula to describe them but also they maximize the number of available states (i.e. entropy), both properties make them very convenient tools for studying $CS$.

By setting to zero a minimum number of $DR^{c}$s: 27 in $x$, 6 in $y$ and 91 in $z$ (1.9\%, 0.4\% and 6.6\% respectively). We obtain a $GCDS$-cone for the whole $\alpha$-carbon chain

\vskip 3mm
$\{\{\ose{ 1} 49 \
     \ose{ 2} 48 \
     \ose{ 3} 29 \
     \ose{ 4} 27 \
              28 \
              30 \
     \ose{ 5} 31 \osd{ 1} \
              52 \osd{ 2} \
              47 \
     \ose{ 6} 32 \
     \ose{ 7} 53 \osd{ 3} \
              50 \osd{ 4} \
     \ose{ 8} 26 \osd{ 5} \
              51 \osd{ 6} \
              21 \
              23 \
              24 \
     \ose{ 9} 19 \
     \ose{10} 20 \
     \ose{11} 25 \
              33 \osd{ 7} \
              46 \
              55 \
     \ose{12} 54 \osd{ 8} \
              22 \osd{ 9} \ \newline
\hspace*{4mm} 18 \
              34 \osd{10} \
              45 \osd{11} \
     \ose{13} 17 \osd{12} \
     \ose{14}  5 \
              44 \
     \ose{15}  8 \
     \ose{16}  6 \osd{13} \
              35 \
              43 \
     \ose{17}  9 \osd{14} \
              16 \
     \ose{18} 11 \osd{15} \
               7 \osd{16} \
              36 \
     \ose{19}  3 \
               4 \
     \ose{20} 10 \osd{17} \
              42 \
     \ose{21} 37 \osd{18} \
     \ose{22} 15 \osd{19} \
     \ose{23} 12 \osd{20} \ \newline
\hspace*{4mm}
     \ose{24} 41 \osd{21} \
              40 \osd{22} \
              38 \
     \ose{25} 14 \osd{23} \
              13 \osd{24} \
              39 \osd{25}\}_{x}$ , \newline
\hspace*{1mm}
  $\{\ose{ 1} 15 \
              16 \
     \ose{ 2} 17 \osd{ 1} \
     \ose{ 3} 14 \osd{ 2} \
     \ose{ 4} 18 \osd{ 3} \
     \ose{ 5} 36 \
     \ose{ 6} 13 \osd{ 4} \
              37 \osd{ 5} \
     \ose{ 7} 19 \
     \ose{ 8} 34 \osd{ 6} \
              12 \
              35 \
              38 \osd{ 7} \
     \ose{ 9} 11 \osd{ 8} \
              20 \
              33 \
     \ose{10} 39 \osd{ 9} \
     \ose{11} 46 \
     \ose{12} 10 \osd{10} \
              32 \
              40 \
              47 \osd{11} \ \newline
\hspace*{4mm}
     \ose{13} 21 \osd{12} \
     \ose{14} 45 \osd{13} \
              44 \
     \ose{15} 31 \
     \ose{16}  9 \
              48 \osd{14} \
              41 \osd{15} \
     \ose{17} 22 \osd{16} \
     \ose{18} 42 \
              49 \
              50 \osd{17} \
               8 \
              30 \
              43 \
              51 \osd{18} \
     \ose{19} 23 \
     \ose{20} 24 \osd{19} \
     \ose{21}  7 \
              52 \
     \ose{22} 29 \
     \ose{23}  4 \
              53 \
              54 \osd{20} \ \newline
\hspace*{4mm}
     \ose{24} 26 \
              27 \osd{21} \
               5 \osd{22} \
               6 \
              25 \
              28 \
              55 \osd{23} \
               3 \osd{24}\}_{y}$ , \newline
\hspace*{1mm}
  $\{\ose{ 1} 26 \
     \ose{ 2} 27 \
     \ose{ 3}  8 \
              10 \
     \ose{ 4}  7 \
     \ose{ 5} 25 \
     \ose{ 6} 11 \
     \ose{ 7} 13 \osd{ 1} \
               9 \osd{ 2} \
               6 \
              24 \
     \ose{ 8} 12 \
     \ose{ 9} 28 \osd{ 3} \
              33 \osd{ 4} \
              31 \osd{ 5} \
     \ose{10} 34 \
     \ose{11} 15 \osd{ 6} \
              32 \
     \ose{12} 29 \
     \ose{13} 17 \osd{ 7} \
     \ose{14} 14 \osd{ 8} \
               5 \
              23 \
     \ose{15}  4 \ \newline
\hspace*{4mm}
              22 \
              35 \
              36 \
              40 \
              41 \
     \ose{16}  3 \osd{ 9} \
              30 \osd{10} \
              39 \osd{11} \
              16 \
              21 \
              43 \osd{12} \
     \ose{17} 38 \osd{13} \
              18 \
              19 \osd{14} \
              20 \osd{15} \
              37 \
              42 \
              44 \osd{16} \
     \ose{18} 55 \osd{17} \
     \ose{19} 45 \
              48 \
     \ose{20} 52 \osd{18} \
              51 \osd{19} \ \newline
\hspace*{4mm}
     \ose{21} 46 \
              47 \osd{20} \
              54 \
     \ose{22} 49 \
              53 \osd{21} \
              50 \osd{22} \}_{z}\}$ \hspace*{86.7mm} (13)

This sequence sets the boundary for the molecular dynamics trajectory in [15] in a compact form\footnote[6]{$\alpha$-carbons from end-residues 1, 2, 56, 57 and 58 are not included because they add disorder, unnecessarily augmenting the volume of the cone without adding much information.}.

\vspace*{4mm}
{\bf 7. Conclusion}

This paper is an outline of a methodology for the exploration of $CS$.

In [1-4] it was assumed that the small local movements of a molecule can be thoroughly sampled in a MDS, and a procedure was devised for building the whole set of structures that result from the combinations of these small movements. The result is a combinatorial structure called the graph of cells, that gives a global view of a molecular system dynamical conformations.

Although the graph of cells can be fitted in a desktop computer file it encodes a huge amount of structures,
the present paper is a first step in solving the problem of managing this great quantity of information. Three issues have been addressed:

\begin{enumerate}
\item we can give bounds that delimit interesting regions (cones) in $CS$,
\item these cones can be decomposed into a set of smaller ones,
\item it is shown that cones in $CS$ can be described by a combinatorial sequence
\end{enumerate}

This last structure, the generalized compact dominance sequence, has embedded in it the whole set of dominance sequences that are in a cone, and can be hierarchically decomposed into a poset structure. On the other hand the graph of cells can be seen as a set of constraints between the $x$, $y$ and $z$ components of the allowed dominance sequences, then the $GCDS$s and the graph of cells complement each other beautifully, since the conformations of the molecular system can obtained by prunning the poset structure from the $GCDS$ with the constraints from the graph of cells. Moreover, $GCDS$s also have a graphical structure where paths and graphical distances between cells (or $3D$-structures) can be determined, and graphical distances between atoms in a $3D$-structure can be enumerated as well. That makes $GCDS$s well suited as the base structures for the development of a combinatorial Hamiltonian in conformational space.

These issues will be further explored in forthcoming works of this series.

\newpage
{\bf References}

\begin{itemize}

\item[[1]]  J. Gabarro-Arpa,
            A central partition of molecular conformational space. I. Basic structures,
            Comp. Biol. and Chem.
            27 (2003) 153-159.

\item[[2]]  J. Gabarro-Arpa,
            A central partition of molecular conformational space. II. Embedding 3D-structures,
            in
           {\it Proceedings of the 26th Annual International Conference of the IEEE EMBS},
           (San Francisco 2004), pp. 3007-3010.

\item[[3]]  J. Gabarro-Arpa,
            Combinatorial determination of the volume spanned by a molecular system in conformational space,
            Lecture Series on Computer and Computational Sciences
            4 (2005) 1778-1781.

\item[[4]]  J. Gabarro-Arpa,
            A central partition of molecular conformational space. III.
            Combinatorial determination of the volume spanned by a molecular system in conformational space,
            Journal of Mathematical Chemistry
            42 (2006) 691-706.

\item[[5]]  P.G. Mezey
            {\it Potential Energy Hypersurfaces},
            (Elsevier, Amsterdam, 1987).

\item[[6]]  D.J. Wales,
            {\it Energy Landscapes},
            (Cambridge University Press, Cambridge, 2003).

\item[[7]]  K. Henzler-Wildman, D. Kern,
            Dynamic personalities of proteins,
            Nature
            450 (2007) 964-972.

\item[[8]]  S. Fomin and N. Reading, Root systems and generalized associahedra,
            math.CO/0505518 (2005).

\item[[9]]  C.R. Shalizi and C. Moore, 
            What is a macrostate? Subjective observations and objective dynamics,
            cond-mat/0303625 (2003).

\item[[10]] A.D. MacKerell Jr., et al.,
            All-Atom empirical potential for molecular modeling and dynamics studies of proteins,
            J. Phys. Chem. B
            102 (1998) 3586-3616.

\item[[11]] W. Wang, O. Donini, C.M. Reyes, P.A. Kollman,
            Biomolecular simulations: recent developments in force fields, simulations of enzyme catalysis, protein-ligand,
            protein-protein, and protein-nucleic acid noncovalent interactions,
            Annu. Rev. Biophys. Biomol. Struct.
            30 (2001) 211-243.

\item[[12]] H.S.M. Coxeter,
            {\it Regular polytopes},
            (Dover Publications Inc., New York, 1973).

\item[[13]] A. Bjorner, M. las Vergnas, B. Sturmfels, N. White,
            {\it Oriented Matroids},
            (Cambridge University Press, Cambridge, UK, sect. 2, 1993).

\item[[14]] M. Marquart, J. Walter, J. Deisenhofer, W. Bode, R. Huber,
            The geometry of the reactive site and of the peptide groups in trypsin,
            trypsinogen and its complexes with inhibitors,
            Acta Crystallogr. Sect. B
            39 (1983) 480-490.

\item[[15]] J. Gabarro-Arpa, R. Revilla,
            Clustering of a molecular dynamics trajectory with a Hamming distance,
            Comp. and Chem.
            24 (2000) 693-698.

\item[[16]] W. Kabsch,
            A discussion of the solution for the best rotation to relate two sets of vectors,
            Acta Cryst.,
            A34 (1978) 827-828.

\end{itemize}

\end{document}